
\magnification=\magstep1
\baselineskip=24 true pt
\hsize=15.0 true cm
\vsize=8.4 true in
\voffset=0.6 true in
\rightline{IP/BBSR/92-92}
\bigskip
\bigskip
\def \sg {\sqrt{-g}}

\def \cav {{\epsilon}^{ijk}}

\def \eq {\eqno}

\centerline {\bf Determination of pseudo-Goldstone boson-photon coupling}
\centerline {\bf by the differential time delay of pulsar signals}
\bigskip
\centerline {\bf  Subhendra Mohanty and  S.~N.~Nayak}
\centerline {\bf Institute of Physics, Bhubaneswar-751005, India.}
\bigskip
\bigskip
\bigskip
\centerline {\bf ABSTRACT}
\noindent Pseudo-Goldstone bosons couple with photons through a P and T
violating interaction of the form
${\cal {L_I}}=g_{a\gamma \gamma}~ a F\tilde F$.
Strong magnetic fields in rotating compact stars induce a non-zero
${\vec E}\cdot{\vec B}$ outside the stellar surface
which acts as a source for the pseudo-scalar field.
Pulsar signals propagating through this pseudo-scalar `hair' suffer a
differential time lag between the left and the right circularly
polarised modes because of the P and T violating pseudoscalar boson-photon
interaction. Determination of this time lag upto a microsecond
accuracy can lead to a measurment of (or rule out)
pseudo-Goldstone boson-photon coupling upto
$g_{a \gamma \gamma}\leq 4.6 \times 10^{-9}~GeV^{-1}$.
\bigskip
\bigskip
\bigskip
\bigskip
\noindent e-mail: mohanty@iopb.ernet.in\hfil\break
\noindent e-mail: sn@iopb.ernet.in\hfil\break
\vfill
\eject

Goldstone bosons arise in particle physics models when
a global symmetry is spontaneously broken.
These include axions [1] which arise in many different
versions of the Peccei Quinn symmetry breaking and may be the solution of
the strong CP problem. Other examples include Majorons [2] which arise
when the lepton number is broken spontaneously and are essential
in order to reconcile possible heavy neutrinos with astrophysical
bounds (a neutrino heavier than $25eV$ and lighter than a few
$GeV$ has to
decay into lighter neutrinos and Majorons in order not to overclose the
universe). Goldstone bosons also arise in superstring theories [3] in the
same multiplet as the graviton and the dilaton and it is possible that
some of the above mentioned Goldstone bosons are string axions.
The coupling of Goldstone bosons with other fields is determined
by the energy scale of the global symmetry breaking. Generically
psedo-scalar boson-photon coupling can be written as
${\cal L}_{a \gamma \gamma}=g_{a \gamma \gamma}~ a F_{\mu \nu}
{\tilde F}^{\mu \nu}$ with $g_{a \gamma \gamma} \sim {C/v}$
where $C$ is model dependent constant and $v$ is the global symmetry
breaking scale.
Strong magnetic fields outside compact rotating stars [4] induce a non-zero
${\vec E}\cdot{\vec B}$ density outside the stellar surface which
acts as a source for the pseudo-scalar field. Pulsar signals propagating
through this pseudo-scalar `hair' suffer a differential time
lag between the left and right circularly polarised modes because of the
Parity and Time-reversal violating pseudo-scalar photon interaction.
We show that for a typical millisecond pulsar the time lag is
$\delta t \simeq 10^{-6}sec$
for a coupling of the order $g_{a\gamma \gamma}\simeq 10^{-9}GeV$.
Measurement of such a time lag between the two circularly polarised modes
of the order of $10^{-6}sec$ have been
reported in literature [5,6,7]. Due to the large
systematic error involved
these numbers should be regarded as an upper bound on $\delta t$
which translates to an upper bound on the pseudo-Goldstone boson-photon
coupling, $g_{a\gamma \gamma}\leq 4.6 \times 10^{-9}GeV^{-1}$.
Our bound is strictly valid for massless pseudo-scalar fields. We have
estimated that this bound continues to hold for pseudo-scalars with
inverse mass greater than the radius of the neutron star i.e. for
$m_a~<~10^{-10}eV$. Hadronic axions have a non-zero mass (induced by
QCD instantons) in the range of $10^{-5} - 10^{-3}eV$, and our
bound does not hold for them.
Previous astrophysical bounds have considered the coupling of axions
to matter in the cooling rates of the Sun, the red giant stars
and the Supernova 1987A [8,9,10].
The best astrophysical bound for the symmetry breaking scale as
obtained from the cooling rate of red giant stars is
$v~< 10^{-8}GeV^{-1}$. Terrestrial experiments have attempted to
produce axions by laser beams in a cavity with magnetic field.
The upper bound on axion-photon coupling from such experiments is
$g_{a\gamma \gamma}~< 2.5\times 10^{-6} GeV^{-1}$ for
$m_a < 10^{-4}eV$ [11].
Experimental searches for axions in
vicinity of the Sun depend on the conversion of axions to X-rays in the
stellar magnetic field.
Based on a survey of the X-ray flux from the Sun,
Lazarus et al [12] give an upper bound of
$g_{a\gamma \gamma}~<~3.6\times 10^{-9}GeV^{-1}$ for $m_a <~0.03eV$.
Our bound for $g_{a\gamma \gamma}$ is of same order as that of
Lazarus et al. In pulsar signals the time lag
between the two circularly polarised modes for the current upper bound
$g_{a\gamma \gamma}\sim 10^{-9}GeV^{-1}$ is $\delta t\sim 10^{-6}sec$.
If it is possible to analyse
the pulse profiles at a resolution higher than $10^{-6}sec$
then it may be possible to actually observe a pulse splitting between the
left and the right polarised modes and thereby determine the value of
$g_{a\gamma \gamma}$. The time lag also depends upon the frequency
$(\delta t \propto \omega^{-2})$ and by analyzing pulsar
signals at frequencies at lower than $100~MHz$ (for which
$\delta t(100~MHz)=10^{-6}sec$) it may be possible to lower the
bounds on $g_{a\gamma \gamma}$ or make a positive measurement of
pseudoscalar-photon coupling.

Interaction of pseudo-Goldstone bosons with photons is described by the
Lagrangian
$$ {\cal L}={1\over 2}({\partial_\mu}a)({\partial^\mu}a)
-{1\over 4}F_{\mu\nu} F^{\mu\nu}
-{1\over 4}g_{a \gamma \gamma}a F_{\mu\nu} {\tilde F}^{\mu\nu}\eq(1)$$
where
${\tilde F}^{\mu\nu}=({1/\sg}) \epsilon^{\mu\nu\rho\sigma} F_{\rho\sigma}$
is the dual of $F_{\mu\nu}$. The equation of motion
for axion field given by (1) is
$$\Big ({1\over \sg}\Big) \partial_{\mu} (\sg \partial^\mu a)
=~-{g_{a \gamma \gamma}\over 4} F_{\mu\nu} {\tilde F^{\mu\nu}}.\eq(2)$$
In terms of the electromagnetic field strengths $E_i \equiv F_{oi}$
and $B^i =\cav F_{jk}$,
 the source term in the right hand side of the equation (2) is
${g _{a \gamma \gamma}}({1/\sg}) ({\vec E}\cdot {\vec B})$.
In order to have an non-trivial axion field configuration outside
the compact stars we must have an external non-zero pseudoscalar
density ${\vec E}\cdot{\vec B}$. In dyonic and Kerr-Newmann family of
blackholes [13] the external ${\vec E}\cdot {\vec B}$ is generated
by the electric/magnetic charges and rotation and acts as a source
for the pseudo-scalar hair in black holes [13].
Pulsars have a strong dipolar magnetic field with field
strengths $B_0 \simeq 10^{12}$ Gauss which is determined by
measuring the frequency of synchrotron X-ray emissions from
their neighbourhood.
In the oblique rotor model [14] for pulsar emission
the dipole moment $\vec\mu$ of a pulsar is not aligned along the
rotation axis $\vec \Omega$ (while in the case of blackhole they
are parallel).
The dipole axis precesses around the spin axis with an angular
frequency $\Omega$ and radiates energy at the rate
${\dot E} = (-2/3) \mid {\ddot {\vec \mu}}\mid = -(1/6) B_0 ^2~R^6$
$\Omega^4 sin^2\alpha \simeq 10^{38}erg~sec^{-1}$ for a typical pulsar.
The angle $\alpha$ can be determined by measuring $\Omega$ and
$\dot \Omega$ for a given pulsar.
In the aligned rotor model [14] the angle $\alpha \sim 0$
(a small misalignment is necessary to produce the pulsation)
and rate of radiation ${\dot E} \sim -B_0 ^2 R^6 \Omega^4$.
The synchrotron radiation from the plasma
accreting along the magnetic poles appears as the pulsed signal to an
observer in the cone swept out by the radio beam as it
precesses about the spin axis.
The instantaneous magnetic field at point ${\vec r~}$ is given by
$${\vec B}={{B_0 R^3}\over r^3} (3 {\hat r}({\hat r}.{\hat \mu})-
{\hat \mu})\eq(3)$$
where $\mid {\vec \mu}\mid =B_0 R^3$ and $B_0$ is the maximum magnetic
field strength at the surface $(r=R)$ of the pulsar. Typically
$B_0\simeq 10^8 - 10^{12}$ Gauss and $R\sim 12$ km.
Magnetic field in a rotating frame appears as an electric field
$${\vec E}=-{\vec \Omega}\times {\vec r}\times {\vec B}.\eq(4)$$
The magnetic field (and consequently the electric field) has a time
dependent part oscillating with frequency $\Omega$ generated by
${\vec \mu}_\perp$, the component of ${\vec \mu}$
perpendicular to $\vec \Omega$
which precesses with angular frequency $\Omega$.
The time independent part of ${\vec B}$ and ${\vec E}$ are generated
by the parallel component ${\vec \mu}_\parallel$ of ${\vec \mu}$.
The time average of field strengths at
time scale $\geq {\Omega^{-1}}$ is determined by
$<{\vec \mu}>={\mu_\parallel}= B_0 R^3 cos\alpha$
(where $\alpha$ is the angle between ${\vec \mu}$ and ${\vec \Omega}$).
The average value of electric field outside the stellar surface is
then determined by (3) and (4) by matching the boundary conditions at $r=R$.
Since ${\vec E}=0$ in the highly conducting interior region, the
component of ${\vec E}$ parallel to the surface is zero at the
boundary and the perpendicular component of ${\vec E}$ changes
discontinuously at $r=R$. The average electric field outside the
stellar surface $(r>R)$ is given by [15]
$$<{\vec E}^{ext}(r)>={{B_0 R^5 \Omega cos\alpha}\over r^4}
\Big [3 (sin^2 \theta -{2\over 3}) {\hat r} - 2 sin\theta
cos\theta~ {\hat \theta}\Big].\eq(5)$$
The average value of the pseudoscalar density ${\vec E}^{ext}\cdot\vec B$ is
$$<{\vec E}^{ext}\cdot {\vec B}>=- {B_0}^2 \Omega cos\alpha ~ {{R^8}\over r^7}
{cos^3}\theta\eq(6)$$
which appears as the source term for the
pseudo-scalar fields in equation (2). In the absence of pseudo-scalar
coupling the strong electric field (5) which is of the order
$E\sim {10^8}$ volts/cm (for $\Omega=1sec^{-1}$ and $B_0 =10^{12}$ Gauss)
will create a space charge of particles with charge density
$\rho=({1/4\pi})~{\vec \nabla}\cdot {\vec E}=
-({1/\pi})~{\vec B}\cdot{\vec \Omega}$
$=({B_0 R^3 \Omega/2\pi r^3})
(sin^2 \theta -2cos^2 \theta).$
This space charge is of the order of $10^{11}$ $cm^{-3}$
 and could effectively screen the electric field.
 Because of the $g_{a \gamma \gamma}
aF{\tilde F}$ coupling however the electric field
is screened by the pseudo-scalar field
and not primarily by the ionised plasma.
This follows from the equation of motion
of electromagnetic fields modified by the pseudo-scalar coupling term
$$D_\mu F^{\mu\nu}=g_{a \gamma \gamma}
(\partial _\mu a)\tilde F^{\mu\nu},\eq(7)$$
which in terms of $\vec E$ and $\vec B$ gives the modified Maxwells equations
$${\vec \nabla}.{\vec E}=- g_{a \gamma \gamma}
({\vec {\nabla a}})\cdot{\vec B},\eq(8a)$$
$$-{\partial_0}{\vec E} +{\vec \nabla}\times {\vec B}=
g_{a \gamma \gamma} \Big [({\vec {\nabla a}})\times {\vec E}
+(\partial_0 a){\vec B}\Big].\eq(8b)$$
The remaining two Maxwells equations follow from the Bianchi
identity $\partial_\mu {\tilde F^{\mu\nu}}=0$ which gives
$${\vec \nabla}\cdot ~{\vec B}=0,\eq(9a)$$
$${\partial_0 {\vec B}}+{\vec \nabla}\times {\vec E}=0.\eq(9b)$$
 From equation (8a) it follows that the effective field experienced by the
plasma is
$$ {\vec E_{eff}} = {\vec E}
 +g_{a\gamma \gamma} a {\vec B}.\eq(10)$$
Therefore a nonzero $a$ and $\vec B$ will anti-screen
the electric field.
The space charge induced in the plasma will neutralise $E_{eff}$ in (10)
but even with the plasma polarisation ${\vec E}$ can remain non-zero.
The standard picture of pulsar
magnetosphere [15] is no longer valid in the presence of pseudo-scalar
fields. This may have other observable consequences
besides the one we study here.

We assume the geometry outside the surface of the pulsar is described
by the Kerr metric
$$ds^2=\Big (1-{2GM\over r}\Big)dt^2 -\Big (1-{2GM\over r}\Big)^{-1}dr^2
-r^2\Big (d\theta^2 +sin^2 \theta d\phi^2\Big)$$
$$-{4GMA\over r}sin^2 \theta d\phi dt.\eq(11)$$
The form (11) is an approximation to the Kerr metric, valid as long as
the angular momentum per unit mass $A~<< GM$. For a typical millisecond
pulsar $(M=~M_{\odot}, R=12~ km)$,
we have $A=10^{-6}$ km and $GM=1.5 km$.
Therefore (11) is an adequate approximation to the Kerr metric.
For the metric given by (11) the equations of motion (2) for the
axisymmetric, time independent solution $a(r,\theta)$ reduces to
$${1\over {r^2}}{\partial\over {\partial r}}
\Big [r^2\Big (1-{2GM\over r}\Big){\partial\over {\partial r}}
a(r,\theta)\Big]+{1\over {r^2 sin ^2 \theta}}{\partial\over
{\partial \theta}}\Big [sin \theta {\partial \over {\partial \theta}}
a(r,\theta)\Big]$$ $$=~{g_{a\gamma \gamma}}
{}~(B_0 ^2 R^8 \Omega ~cos\alpha)~{1\over {r^7}}~cos^3 \theta.
\eq(12)$$
The Greens function which is solution of $\nabla^2 G(x,y)=-
{\delta ^3 (x-y)/\sg}$ for the Laplacian operator in (12) is
given by [13]
$$G(r,\theta,r_0 ,\theta_0 )
=\Sigma_{l=0}^{\infty}~C_lP_l
\Big ({r_0\over GM}-1\Big) Q_l\Big ({r\over GM}-1\Big)
P_l(\lambda),~~~r>r_0$$
$$~~~~~~~~~~~~~~~~~~~~~~~=\Sigma_{l=0}^{\infty}~C_lQ_l
\Big ({r_0\over GM}-1\Big)P_l\Big ({r\over GM}-1\big)
P_l(\lambda),~~~r<r_0, \eq(13)$$
where $\lambda=cos \theta cos \theta_0 +sin \theta sin \theta_0$
, $C_l={(2l-1)/ 4\pi M}$ and $P_l$ and $Q_l$ are the
two associated Legendre functions.
The solution of the inhomogenous equation (12) is
$$a(r,\theta)=~-{\int_R ^r}dr_0 {\int_0 ^\pi}d\theta_0
{\int_0 ^{2\pi}}d\phi_0 {r_0 ^2} sin^2 \theta_0~
G(r,\theta,r_0 ,\theta_0 )$$
$$\times \Big({{g_{a \gamma \gamma}}}
 {B_0 ^2} R^8 cos\alpha~ \Omega\Big)~
{{cos^3 \theta_0}\over {r_0 ^7}}.\eq(14)$$
Substituting (13) in (14) and evaluating the integral we obtain
the Goldstone boson configuration outside pulsar $(r>R)$
$$a(r,\theta)= -\Big({2\over 575}\Big)~
{{g_{a \gamma \gamma}~ {B_0 ^2}~ R^8~ \Omega~cos\alpha}\over {(GM)^3}}
{}~{cos\theta\over r^2} + O\Big({1\over r^3}\Big).\eq(15)$$
Propagation of electromagnetic radiation through this coherent
pseudo-Goldstone boson field is governed by the wave equation
obtained by combining the Maxwells equations (8) and (9)
$${\nabla_\mu}{\nabla^\mu}{\vec B}=-~g_{a \gamma \gamma}
(\vec {\nabla a}) \times \partial_0 {\vec B}.\eq(16)$$
With the eikonal ansatz
$${\vec B} (x,t)={\vec {\cal B}}~e^{iS(x,t)}\eq(17a)$$
and $$-i\partial_\mu {\vec B} = (\partial_\mu S){\vec B}
 = k_\mu {\vec B} \eq(17b)$$
the wave equation reduces to
$${k_\mu} {k^\mu}{\vec {\cal B}}= i~g_{a\gamma \gamma}
({\vec {\nabla a}}\times \omega {\vec {\cal B}})\eq(18)$$
where $k^\mu=~(\omega ,{\vec k})$ is the photon four momentum
and ${\vec {\cal B}}$ is the magnetic field of the radiation.
The pulsed radiation
from pulsars originates from the polar regions of the pulsar [16].
In this region $a\simeq a_0 (1/{r^2})$ and
${\vec {\nabla a}} = -({2a_0}/r^3)~ \hat r$ is radial.
Consider right and left circularly polarised
combinations for the transverse magnetic field
${\vec {\cal B}}$
$$ {\vec {\cal B}}_\pm =
({\vec {\cal B}_i} \pm i{\vec {\cal B}}_j)\eq(19)$$
where ${\vec {\cal B}}_i $ and ${\vec {\cal B}}_j $
are the components of ${\vec {\cal B}}$
along any two fixed directions $\hat e_i$ and $\hat e_j$
orthogonal to $\hat r$.
Substituting (19) in (18) we find that the equations for the two
circularly polarised modes ${\vec {\cal B}}_\pm$ decouple as
$$k_\mu k^\mu {\vec {\cal B}}_\pm~ \mp~ g_{a \gamma \gamma}
({\partial_r a})~\omega~
{\vec {\cal B}}_\pm =0.\eq(20)$$
This shows that the propagating modes in a background with non-zero
pseudoscalar gradients are circularly polarised modes [17] with
different phase velocities.
The dispersion relation for the ${\vec {\cal B}}_\pm$ modes propagating
radially from the poles is
$${\omega^2}\Big(1-{2GM\over r}\Big)^{-1}-k_r^2
 \Big(1-{2GM\over r}\Big) =\pm
g_{a \gamma \gamma}(\partial_r a)k_0 .\eq(21)$$
The Eikonal phase can be written as
$$S(r,t)= \omega t-{\int^r} k_r dr.\eq(22)$$
The propagation time $t$ to reach a point $r$ is given by the orbit
equation
${{\partial S}/{\partial \omega}}=0$ which gives
$$t={\partial\over {\partial \omega}}{\int_R ^r} k_r dr.\eq(23)$$
Substituting for $k_r$ from (20) we have
$$t_\pm ={\int_R ^r}dr~\Big (1+{2GM\over r}\Big )
\mp g_{a\gamma \gamma}^{2}
{\int_R ^r}dr~ {{(\partial _r a)^2}\over {4 \omega^2}}\Big (1-{2GM\over r}
\Big ), \eq (24)$$
where $t_{+}(t_-)$ are the arrival times of the $+(-)$
circularly polarised modes. Substituting for $\partial_r a(r)$
from (15) (with $cos \theta=1$) and evaluating the integrals in (24)
we obtain the arrival times of the
pulses from a pulsar at distance r from the earth
$$t_{\pm}=r+2GM ln\Big({r\over R}\Big)\pm {1\over 5}\Big({2\over 575}
\Big )^2~{{g_{a\gamma \gamma}^4~B_0^4~R^{11}~\Omega ^2~cos^2 \alpha}
\over {\omega^2~(GM)^6}}. \eq(25)$$
The second term is the familiar time delay due to the gravitational
field of the pulsar and the last term shows the birefrigence
properties of the pseudo-scalar background.
The differential time delay between the positive and the
 negative polarised modes is
$$\delta t=t_+ - t_-
={2\over 5}\Big ({2\over 575}\Big )^2 {{g_{a\gamma\gamma}^4
{}~B_0^4 ~R^{11}~\Omega ^2~ cos^2 \alpha}\over {\omega^2~ (GM)^6}}.
\eq(26)$$
Pulsar signals are recorded in the radio frequency range as
low as $\omega \simeq 100 MHz$. For a typical millisecond pulsar $M=M_{\odot}
=10^{57} GeV$, $B_0=10^{12} Gauss \simeq 10^{-9} GeV^2$, $R=12~ km
=6\times 10^{19} GeV^{-1}$, $cos\alpha\simeq 1$
 the differential time delay turns out
to be $10^{-6}$ sec for $g_{a\gamma \gamma}\simeq 10^{-9}GeV^{-1}$.
By putting an experimental bound on $\delta t$ one can therefore
obtain the following upper bound on the pseudo-Goldstone boson-photon
coupling.
$$g_{a\gamma \gamma}
=~4.6\times 10^{-9}GeV^{-1}~ \Big ({{\delta t}\over {10^{-6} sec}}
\Big )^{1/4}\Big ({ {10^{12} G}\over B_0}\Big )
\Big({{12km}\over R}\Big)^{11/4}
\Big ({{10^{-3}sec}\over \Omega}\Big)
^{1/2}$$
$$\times \Big ({{\omega}\over 100MHz}\Big )^{1/2}
 \Big ({M\over {M_\odot}}
\Big)^{3/2}. \eq(27)$$
Loseco et al [5] have analysed the signals from the millisecond
pulsar PSR \hfil \break 1937+214 using the data of
 Cordes and Stinebring [6].
They report that a time lag $5\times 10^{-6}$ sec is observed
between the positive and the negative polarised modes.
Klien and Thorsett [7] report a differential time lag of
$0.37\pm 0.67 \mu sec$.
At the one sigma limit one can put an upper bound on the time
lag to be $\delta t \leq 10^{-6}$ sec. This leads to the upper
bound on the goldstone boson photon coupling
$$g_{a\gamma \gamma} \leq 4.6\times 10^{-9} GeV^{-1}. \eq(28)$$

 From (26) it is clear that by going for lower frequency pulses one
should observe a larger time lag
since $\delta t \propto \omega^{-2}$. Therefore by analysing lower
frequency signals it should be possible to observe a non-zero time
lag which would be a positive signature of pseudo-scalar photon
interaction.
\vfill
\eject
\centerline{\bf References}
\item {1} R. D. Peccei and H. R. Quinn, Phys. Rev. Lett.
{\bf 38} (1977) 1440; Phys. Rev. {\bf D16} (1977) 1791.

\item {2} Y. Chikashige, R. N. Mohapatra and R. D. Peccei,
Phys. Rev. Lett. {\bf 45} (1980) 1926; Phys. Lett {\bf B 98} (1981) 265;
\item { } G. B. Gelmini and M. Roncadelli, Phys. Lett {\bf B 99} (1981) 411;
\item { } H. Georgi, S. L. Glashow and S. Nussikov, Nucl. Phys. {\bf B 193}
(1981) 297.

\item {3} M. Green, J. Schwarz and E. Witten, `Superstring theory'
Cambridge U.P; Cambridge, 1987.

\item {4} S. L. Shapiro and S. A. Teukolsky, `Black holes, White dwarfs,
and Neutron stars', John Wiley and Sons, (1983).

\item {5} J. M. Loseco, G. E. A. Matsas, A. A. Natale and J. A. F. Pacheco,
Phys. Lett {\bf A 138} (1989) 5.

\item {6} J. M. Cordes and D. R. Stinebring, Ap J. Lett. {\bf 277} (1984) 53.

\item {7} J. R. Klien and S. E. Thorsett, Phys. Lett {\bf A 145} (1990) 79.

\item {8} G. G. Raffelt, Phys Rep. {\bf 198} (1990) 1.

\item {9} M. S. Turner, Phys Rep. {\bf 197} (1990) 67.

\item {10} E. W. Kolb and M. S. Turner, `The Early Universe',
Addison-Wesley, (1990).

\item {11} Y. Semertzidis et al, Phys. Rev. Lett.
{\bf 25} (1990) 2988.

\item {12} D. M. Lazarus et al, Phys. Rev. Lett.
{\bf 69} (1992) 2333.

\item {13} B. Campbell, N. Kaloper and K. A. Olive, Phys. Lett {\bf B 263}
(1991) 364; Phys. Lett {\bf B 285} (1992) 199.

\item {14} F. Pacini, Nature {\bf 216} (1967) 567;
 Nature {\bf 219} (1968) 145;
\item {  } J. E. Gunn and J. P. Ostriker, Nature {\bf 221} (1968) 254.

\item {15} P. Goldreich and W. H. Julian, Ap. J. {\bf 157} (1969) 869.

\item {16} V. Radhakrisnan and D. J. Cooke,
Astrophys. Lett. {\bf 3} (1969) 225.

\item {17} Sean M. Caroll, George B. Field and Roman Jackiw, Phys. Rev.
 {\bf D41} (1990) 1231;
\item { } Sean M. Caroll and George B. Field, Phys. Rev.
 {\bf D43} (1991) 3789.
\vfill
\end